\documentclass[
    reprint,
    aps,
    superscriptaddress,
    nobalancelastpage,
    floatfix
]{revtex4-2}
\usepackage[english]{babel}
\usepackage{hyperref}

\usepackage{amsmath}
\usepackage{amssymb}
\usepackage{amsfonts}
\usepackage{bm}
\usepackage{physics}
\usepackage{siunitx}
\ExplSyntaxOn
\msg_redirect_name:nnn { siunitx } { physics-pkg } { none }
\ExplSyntaxOff

\usepackage{graphicx}
\usepackage{xcolor}
\usepackage{tikz}
\usepackage{booktabs}
\usepackage[caption=false]{subfig}

\newcommand{\Renyi}{R\'enyi}
\newcommand{\Neel}{N\'eel}

\newcommand{\onehalf}{\frac{1}{2}} 
\newcommand{\MinEnt}{S_{\infty}}
\newcommand{\even}{\text{(e)}}
\newcommand{\odd}{\text{(o)}}
\newcommand{\gs}{\text{gs}}
\newcommand{\pmax}{p{\smash[t]{\strut}}_{\text{max}}}
\newcommand{\imax}{i_{\text{max}}}
\newcommand{\gseven}{\gs^{\even}}
\newcommand{\gsodd}{\gs^{\odd}}
\newcommand{\neeleven}{\text{\Neel{}}^{\even}}
\newcommand{\neelodd}{\text{\Neel{}}^{\odd}}

\newcommand{\IFT}{Instituto de Física Teórica, UAM/CSIC, Universidad Autónoma de Madrid, Madrid, Spain}

\newcommand{\DIPC}{DIPC - Donostia International Physics Center, Paseo Manuel de Lardiz{\'a}bal 4, 20018 San Sebasti{\'a}n, Spain}
\newcommand{\UPV}{EHU Quantum Center and Department of Physical Chemistry, University of the Basque Country UPV/EHU, P.O. Box 644, 48080 Bilbao, Spain}
\newcommand{\CERN}{European Organization for Nuclear Research (CERN), Geneva 1211, Switzerland}
\newcommand{\IKERBASQUE}{IKERBASQUE, Basque Foundation for Science, Plaza Euskadi 5, 48009 Bilbao, Spain}
\newcommand{\PHYSCHEM}{Department of Physical Chemistry, University of the Basque Country UPV/EHU, Box 644, 48080 Bilbao, Spain}

\renewcommand{\paragraph}[1]{\bigbreak\emph{#1} ---}

\begin{document}

\title{Oddities in the Entanglement Scaling of the Quantum Six-Vertex Model}

\author{Sunny Pradhan}
\email{sunny.pradhan@ehu.eus}
\affiliation{\UPV}

\author{Jes\'us Cobos}
\affiliation{\UPV}
\affiliation{\PHYSCHEM}

\author{Enrique Rico}
\affiliation{\UPV}
\affiliation{\DIPC}
\affiliation{\IKERBASQUE}
\affiliation{\CERN}

\author{Germ\'an Sierra}
\affiliation{\IFT}

\begin{abstract}
We investigate the entanglement properties of the Quantum Six-Vertex Model on a cylinder, focusing on the Shannon-R\'enyi entropy in the limit of R\'enyi order $n = \infty$.
This entropy, calculated from the ground state amplitudes of the equivalent XXZ spin-1/2 chain, allows us to determine the Renyi entanglement entropy of the corresponding Rokhsar-Kivelson wavefunctions, which describe the ground states of certain conformal quantum critical points.
Our analysis reveals a novel logarithmic correction to the expected entanglement scaling when the system size is odd.
This anomaly arises from the geometric frustration of spin configurations imposed by periodic boundary conditions on odd-sized chains.
We demonstrate that the scaling prefactor of this logarithmic term is directly related to the compactification radius of the low-energy bosonic field theory description, or equivalently, the Luttinger parameter.
Thus, this correction directly probes the underlying Conformal Field Theory (CFT) describing the critical point.
Our findings highlight the crucial role of system size parity in determining this model's entanglement properties and offer insights into the interplay between geometry, frustration, and criticality.
\end{abstract}

\maketitle


Investigating extended quantum systems has always been an important line of research in condensed matter physics.
An essential instrument in this regard is the study of the entanglement in the ground state, whose behavior can reveal some non-trivial features about a quantum system \cite{amico2008entanglement}.
Just to give some examples, through entanglement one can measure the central charge in massless systems in $1+1$ dimensions \cite{holzhey1994entropy, vidal2003entanglement, calabrese2004entanglement} or detect topological phases and measure the quantum dimensions of anyonic excitations in $2 + 1$ dimensions \cite{hamma2005bipartite,levin2006detecting,kitaev2006topological}.

In the case of two-dimensional quantum critical systems, in particular, \emph{conformal quantum critical points} \cite{ardonne2004topological}, besides the expected area law for entanglement, \emph{subleading terms} can also appear.
\emph{Logarithmic terms} that grow as $\log L$ can appear, with an overall factor that depends on the central charge, topology, smoothness of the manifold \cite{fradkin2006entanglement}, and also the boundary conditions \cite{zaletel2011entanglement}.
Moreover, also the constant terms are rather interesting \cite{hsu2009entanglement, stephan2009shannon, oshikawa2010cft}, as they depend on Affleck-Ludwig boundary entropy \cite{affleck1991groundstate}.
The main takeaway message here is the following: at criticality, the entanglement is largely determined by the system's long-range behavior, which can be modeled via an appropriate conformal field theory (CFT) \cite{difrancesco2012conformal, ginsparg1988cft}.
Therefore, the entanglement scaling can be used in different situations as a tool to extract physical data on the underlying CFT.

In this work, we interrogated ourselves about one particular issue: \emph{does the system size parity have any effect on entanglement?} If yes, \emph{does it reveal anything interesting}?
Note that by system size parity we simply mean that the size is even or odd.
In the case of one-dimensional critical systems, the answer is affirmative.
For example, in \cite{calabrese2010parity, fagotti2011parity} the parity of the system size can lead to oscillations in the entanglement that depend on the Luttinger parameter $K$ and the Fermi momentum $k_F$.
Moreover, in \cite{giampaolo2019frustration, torre2024topological} it is even suggested that the frustration introduced by an odd length size and antiferromagnetic couplings on a periodic chain can drastically change the phase diagram of a spin chain.

However, not much is known about the effects of parity (\emph{oddities}) in two-dimensional systems \cite{ju2012entanglement,stephan2013entanglement}.
For this reason, we focus on conformal quantum critical points, whose ground state can be written as \emph{generalized Rokhsar-Kivelson (RK) wavefunction} \cite{rokhsar1988dimer, ardonne2004topological, henley2004rkpoints}.
This type of state was first introduced for quantum dimer models \cite{rokhsar1988dimer}.
It is a  superposition of all the microscopical configurations $\{\ket{c}\}$ that satisfy a certain set of local constraints with Boltzmann weigths $e^{-E(c)/2}$:
\begin{equation}
    \ket*{\text{RK}} = \frac{1}{\sqrt{\mathcal{Z}}} \sum_{c} e^{- E(c)/2} \ket{c},
\end{equation}
where the normalization is given by the partition function $\mathcal{Z} = \sum_{c} \exp(- E(c))$.
These wavefunctions are quite common when discussing topological order, like in the case of the toric code \cite{kitaev2003toric} or quantum spin liquids \cite{savary2016spinliquids}.

\begin{figure*}[t]
    \centering
    \hspace{0.5cm}
    \subfloat[][\label{fig:cylinder}]{\includegraphics{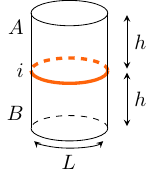}}
    \hfill
    \subfloat[][\label{fig:partition}]{\includegraphics{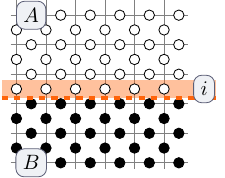}}
    \hfill
    \subfloat[][\label{fig:six_vertex}]{\includegraphics{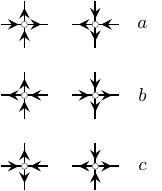}}
    \hfill
    \subfloat[][\label{fig:neel_state}]{\includegraphics{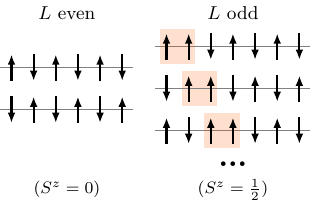}}
    \hspace{0.5cm}

    \caption{(a) The 2D quantum system lives on a cylinder of circumference $L$ and height $2h$; it is bipartite into two subsystems each of height $h$.
    (b) With degrees of freedom on the links and local constraints around each vertex, the boundary configurations $i$ enumerate the Schmidt eigenvalues and eigenvectors of the decomposition of the RK wavefunction; this is possible because by fixing $i$, we also constraint the possible configurations of $A$ and $B$, which allows writing the Schmidt eigenvectors again in an RK form.
    (c) Allowed vertex configurations of the 6VM with their weights;
    (d) The different maximal states (\Neel{} states) when $L$ is even or odd.
    In the even case (\emph{left}) there are only two possibilities independent of system size; in the odd case (\emph{right}) the total spin is $S^z = \pm 1/2$ and there is always a pair of parallel spins.
    It can be in $L$ possible positions, leading to an extensive degeneracy in the maximal state when $L$ is odd.
    }
\end{figure*}

In this paper we consider the case where the RK wavefunction is built from the configurations of the \emph{Six-Vertex Model} on an infinite cylinder.
We show that when the size of the diameter is odd an extra logarithmic term appears in the entanglement scaling, with a factor that depends on the underlying CFT, namely the Luttinger parameter $K$.
The physical origin of this term is traced back to an extra spin-$\onehalf$ excitation living on the border between the two partitions of the cylinder.
The border can effectively be described as a Luttinger liquid and the log term coefficient depends on the power-law behavior of the spin-spin correlators, which is governed by the Luttinger parameter.

To probe the entanglement we employ the technique developed in \cite{stephan2009shannon}, which is perfectly tailored for this situation, where we have a quantum system on a square cylindrical lattice, spin-$\onehalf$ degrees of freedom and local constraints around each vertex (see Fig.~\ref{fig:cylinder}).
The authors show that the Schmidt eigenvalues $p_i$ of the RK wavefunction---enumerated by the boundary configurations, as discussed in Fig.~\ref{fig:partition}---are equivalents to the probabilities of the ground state of a 1D quantum system in a given basis.
To be more specific, $p_i = \abs{ \braket{i}{\gs} }^2$ in the limit $h \to \infty$ of infinite cylinder.
The basis $\{\ket{i}\}$ is the one defined by the transfer matrix of the classical model, while $\ket{\gs}$ is the ground state of the 1D quantum Hamiltonian obtained from it.

The $n$-th \Renyi{} entanglement entropy of a state $\rho$ over a bipartition $A \cup B$ is defined as
\begin{equation}
    S_n = \frac{1}{1 - n} \log \Tr \rho_A^n = \frac{1}{1 - n} \log \sum_{i} p_i^n
    \label{eq:renyi_ent_def}
\end{equation}
where $\rho_A = \Tr_B \rho$ is the reduced density matrix of $A$ and the $p_i$'s are the Schmidt eigenvalues \cite{nielsen2010quantum}.
Using the technique described in \cite{stephan2009shannon}, the computation of the \Renyi{} entanglement entropy of the 2D system is equivalent to the computation of the \emph{Shannon-\Renyi{} entropy} \cite{luitz2004spetroscopy} of the 1D quantum system.
The latter is defined with the same expression at the right of \eqref{eq:renyi_ent_def} but with $p_i = \abs{\braket{i}{\gs}}^2$ instead of the Schmidt eigenvalues.
We are particularly interested in the limit $n = \infty$, where
\begin{equation}
    \MinEnt = - \log \pmax
    \quad \text{with} \quad
    \pmax = \max_i \abs{\braket{i}{\gs}}^2
    \label{eq:def_min_entropy}
\end{equation}
From the entanglement point of view, $\MinEnt$ corresponds to \emph{single-copy entanglement} \cite{eisert2005singlecopy, peschel2005singlecopy}, while from the Shannon information point of view $\MinEnt$ is known as the \emph{min-entropy} \cite{renner2004renyi, konig2009entropy}.
In this context, it can be used to quantify the spread of the ground state on the basis $\{\ket{i}\}$ \cite{stephan2009shannon}.
In this manuscript we mainly develop theoretical arguments that show the emergence of a logarithmic correction to the min-entropy $\MinEnt$ in the case of an odd chain.

The min-entropy $\MinEnt$ plays a special role in the study of the \Renyi{} entanglement entropy in critical RK wavefunctions, or, equivalently, Shannon-\Renyi{} entropy of critical chains.
It has been proven that the subleading constant of $S_n$, as a function of the Renyi order $n$, goes through a phase transition, in the sense that the dependence on $n$ abruptly changes while crossing a critical index $n_c$ \cite{stephan2011phase, stephan2010renyi}.
In fact, for $n > n_c$ we are in a ``locked phase'' where only the states with largest probability contributes to the entropy $S_n$.
Given that $\MinEnt$ depends only on the largest probability $p_{\text{max}}$, its behavior influences all the other $S_n$'s in the locked phase.
It is natural, then, to ask if the results about the min-entropy $\MinEnt$ extends to all the other entropies $S_n$ with $n \neq \infty$.
For this reason, numerical computations of $S_{n}$ with exact diagonalization have been carried out, later in the manuscript.
The results indeed show that logarithimic corrections appear also for $n \neq \infty$.
Unfortunately, we were not able to establish if these corrections had any universal character to them


\paragraph{Model}
We consider an RK wavefunction built from the states of a \emph{Six-Vertex Model} (6VM) \cite{baxter1982exactlysm}.
It is a statistical model on a square lattice with binary degrees of freedom on the links, pictured as arrows, where the number of incoming and outgoing arrows at each vertex has to match.
The allowed vertex configurations and their weights are shown in Fig.~\ref{fig:six_vertex}.
For the 6VM, the equivalent 1D quantum Hamiltonian is the \emph{XXZ spin-$1/2$ chain}:
\begin{equation}
    H_{\text{XXZ}} = \sum_{i=1}^{L} \qty(
    \sigma^x_i \sigma^x_{i+1} +
    \sigma^y_i \sigma^y_{i+1} +
    \Delta \sigma^z_i \sigma^z_{i+1} ),
    \label{eq:XXZ_hamiltonian}
\end{equation}
a well-known integrable model.
The \emph{anisotropy parameter} $\Delta$ is a function of the 6VM weights via the expression $\Delta = (a^2 + b^2 - c^2)/(2ab)$.
The natural basis coming from the transfer matrix is the computational basis, where $\sigma^z$ is diagonal, and the cylindrical geometry in Fig.~\ref{fig:cylinder} means that we have to impose periodic boundary conditions (PBC) on the spin chain.
We briefly review the phase diagram of the XXZ chain, which consists of three regions \cite{baxter1982exactlysm, cabra2004magnetism}: a ferromagnetic phase for $\Delta \leq -1$, an antiferromagnetic phase for $\Delta > 1$, and a whole \emph{critical phase} for $-1 < \Delta \leq 1$ which will be the focus of this work.
The point $\Delta = 1$ is included in the critical region of the spin chain and it is a phase transition of BKT type \cite{kosterlitz1974xymodel}.
We remark that in the critical phase, at low energy, the model can be described as a massless compactified boson field, which is a CFT with central charge $c = 1$, with compactification radius \cite{eggert1992heisenberg, affleck1988fields}
\begin{equation}
    R = \sqrt{ \frac{1}{2 \pi} - \frac{1}{2 \pi^2} \cos^{-1} \Delta }.
    \label{eq:compactification_radius}
\end{equation}
Equivalently, it can be described as a Luttinger liquid with Luttinger parameter $K = 1/(4 \pi R^2)$ \cite{cabra2004magnetism, giamarchi2003onedim}.

The terminology ``quantum Six-Vertex Model'' is derived from Ref.~\cite{ardonne2004topological},
where it is shown how to obtain an Hamiltonian such that its ground state is an RK wavefunction over the 6VM configurations.
This particular Hamiltonian can also be seen as a $U(1)$ lattice gauge theory, where the local constraints corresponds to a discretized version of Gauss's Law.


\paragraph{Odd number of spins}
To compute $\MinEnt$ we have to know the state $\ket{\imax}$ in the computational basis that maximizes the probability $p_i = \abs{\braket{i}{\text{gs}}}^2$.
In the case of the critical XXZ chain, it is established that such state is the \Neel{} state, i.e.~alternating spin configuration \cite{stephan2009shannon, luitz2004spetroscopy}.
At this point, we have to start to distinguish the cases based on the parity of $L$.
When $L$ is even, each pair of neighboring spins is antiparallel, meaning that the total magnetization along the $z$-axis is zero, $S^z = 0$.
The degeneracy of the maximal state is only two-fold because there are only two distinct \Neel{} states and they differ by a single site shift.

The situation changes drastically when $L$ is odd with PBC.
Not all neighboring spin pairs can be antiparallel, as there is necessarily a \emph{pair of parallel spins} which can be thought of as a \emph{spinon}.
This means that the total spin $S^z$ is either $+1/2$ or $-1/2$ and we have extensive degeneracy (it grows with $L$) due to all possible positions of the parallel pair, the moving spinon.
The difference between $L$ even and $L$ odd is illustrated in Fig.~\ref{fig:neel_state}.

It is easy to imagine that the degeneracy in the maximal state would affect $\MinEnt$ as a function of $L$.
To illustrate this point, consider a wavefunction $\ket{\phi}$ that is an equal superposition of $n$ different states $\ket{j}$, i.e.~$\ket{\phi} = n^{-1/2} \sum_{j=1}^n \ket{j}$.
Then, we have simply $\MinEnt = \log n$, that is logarithmic in $n$, the ``degeneracy''.
This gives us a reason to think that a logarithmic term would likely appear in the min-entropy scaling when $L$ is odd, as a result of the maximal state degeneracy, which signals a moving spinon.
As a start, we can check the presence of any log term in a couple of special points, $\Delta = 1/2$ and $\Delta = 0$.
Remark on notation:
anytime we deal with a quantity $\mathcal{O}(L, N)$ that depends on the length $L$ and the number of particles (spin ups) $N$, we will use the shorthands $\mathcal{O}^{\even}(N)$ and $\mathcal{O}^{\odd}(N)$ for $\mathcal{O}(2N, N)$ and $\mathcal{O}(2N + 1, N)$, respectively.

\begin{figure*}[t]
    \hspace{0.25cm}
    \subfloat[][\label{fig:minent_delta_half}]{\includegraphics[height=4.5cm]{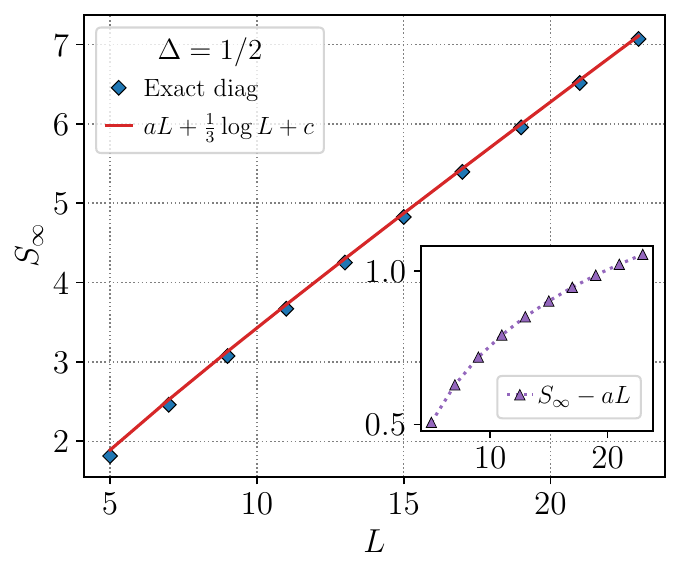}}
    \hfill
    \subfloat[][\label{fig:minent_delta_zero}]{\includegraphics[height=4.5cm]{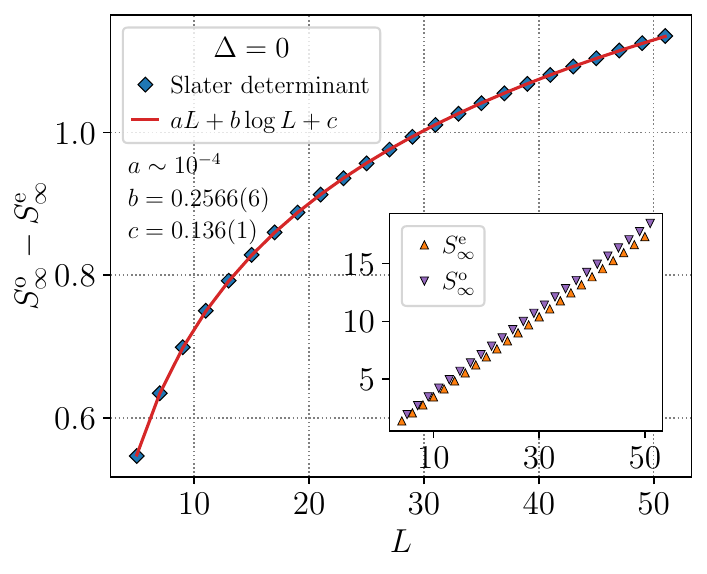}}
    \hfill
    \subfloat[][\label{fig:log_coefficient}]{\includegraphics[height=4.5cm]{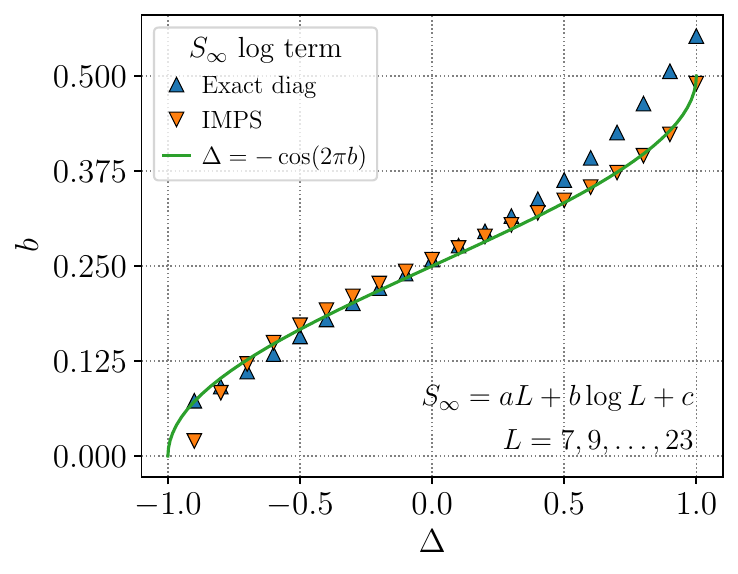}}
    \hspace{0.25cm}
    \caption{(a) The min-entropy $\MinEnt$ at $\Delta = 1/2$ for $L = 5, 7, \dots, 23$ obtained from exact diagonalization.
    The solid line is the analytical expression \eqref{eq:minent_delta_half} derived from \cite{razumov2001spin}.
    The full expression \eqref{eq:minent_delta_half_theor} can be found in Appendix~\ref{app:computation_of_minent_at_delta_half}.
    \emph{Inset.} The plot of min-entropy minus the linear term shows a clear $\log$ behavior.
    (b) The entropy difference $\MinEnt^{\odd} - \MinEnt^{\even}$ at $\Delta = 0$ for $L = 5, 7, \dots, 51$, computed numerically from the Slater determinant \eqref{eq:slater_vandermonde_det}.
    The fit against the function $a L + b \log L + c$ (with $L$ odd) shows $a \sim 10^{-4}$ (negligible), $b = \num{0.2566+-0.0006}$ (very close to $1/4$) and $c = \num{0.136+-0.001}$.
    \emph{Inset.} The entropies $\MinEnt^{\odd}$ and $\MinEnt^{\even}$ are shown separately in the same range of $L$.
    (c) The log term coefficient $b$ of $\MinEnt$ for $\Delta \in (-1, 1]$ (20 points).
    $\MinEnt$ has been computed for $L = 7, 9, \dots, 23$ with exact diagonalization and then fitted against $a L + b \log L + c$.
    The fit errors on $b$ are of order \num{e-3}--\num{e-4}.
    As a comparison, the log coefficient obtained in the same manner from the iMPS Ansatz \eqref{eq:IMPS} (same $L$s) and the theoretical curve $\Delta = - \cos(2 \pi b)$ have also been plotted.
    }
\end{figure*}

\paragraph{Special points}
At $\Delta = 1/2$ we take advantage of the conjectures put forward in \cite{razumov2001spin} (and partially proven in \cite{razumov2007spin}) for the XXZ spin chain with PBC and odd $L$
\footnote{In Ref.~\cite{razumov2001spin} the XXZ Hamiltonian has a global minus, thus the $\Delta = 1/2$ in \eqref{eq:XXZ_hamiltonian} (which have a global plus sign) corresponds to $\Delta = -1/2$ in \cite{razumov2001spin}.
When $L$ is even it is possible to invert the sign in the $xy$ plane with a simple unitary transformation $S^{x/y}_j \mapsto (-1)^{j} S^{x/y}_j$.
This is not so immediate when $L$ is odd, the above mapping leads to extra boundary terms.
Nonetheless, we have verified numerically (with exact diagonalization) that the main statements in \cite{razumov2001spin} still hold even with a global plus sign and $\Delta = 1/2$.
}.
From their findings, we can directly compute $\MinEnt(L)$ and indeed we encounter a $\log$ term:
\begin{equation}
    \MinEnt(L)_{\Delta = \onehalf} = a_{\onehalf} L + \frac{1}{3} \log L + c_{\onehalf},
    \label{eq:minent_delta_half}
\end{equation}
where $a_{\onehalf}$ and $c_{\onehalf}$ are some constants specific to $\Delta = 1/2$.
The full derivation can be found in Appendix~\ref{app:computation_of_minent_at_delta_half}.
In order to verify \eqref{eq:minent_delta_half} we have computed $\MinEnt$ at $\Delta = 1/2$ with exact diagonalization for $L = 5, 7, \dots, 21$.
The results are shown in Fig.~\ref{fig:minent_delta_half} and agree with \eqref{eq:minent_delta_half}.

At $\Delta = 0$ the XXZ Hamiltonian \eqref{eq:XXZ_hamiltonian} can be mapped into a chain of spinless free fermions using a Jordan-Wigner transformation \cite{lieb1961models}.
With PBC, the Hamiltonian is diagonalizable in momentum space and the ground state $\ket{\gs}_{\Delta = 0}$ is obtained by simply filling all the momenta with negative energy.
With $N$ occupied levels only the overlaps that involve exactly $N$ fermions are non-zero and they can be computed via \emph{Slater determinant}.
Indeed, if $1 \leq n_1 < \dots < n_N \leq L$ are the positions of the occupied sites (spin ups), then the overlap
$\Psi(n_1 \dots n_N)_{\Delta = 0} = \braket{n_1 \dots n_N}{\gs(L, N)}_{\Delta = 0}$ is given by the determinant of
\begin{equation}
    \mathsf{M}(n_1 \dots n_N) = \qty( \frac{e^{i q_j n_k}}{\sqrt{L}} )^N_{j,k=1},
\end{equation}
where the $q_j$'s are the occupied momenta and $L$ is the system size.
Given that the occupied momenta are equally spaced, i.e.~$q_j = q_{\text{F}} + 2 \pi j / L$ with $q_{\text{F}}$ being the Fermi momentum, $\det \mathsf{M}$ can be put in a Vandermonde determinant form which yields
\begin{equation}
    \abs{\Psi_{\Delta = 0}(n_1 \dots n_N)}^2 =
    \frac{1}{L^{N}}
    \prod_{i > j}^N \abs{
        e^{i 2 \pi n_i / L} - e^{i 2 \pi n_j / L}
    }^2.
    \label{eq:slater_vandermonde_det}
\end{equation}
For $L$ even we have $L = 2N$, while for $L$ odd we choose $L = 2N + 1$.
In both cases, the maximal state is given by the configuration $n_j = 2j - 1$ (where $j = 1, \dots, N$), up to translational invariance.
The entropy $\MinEnt^{\even}$ in the even case can be evaluated exactly from \eqref{eq:slater_vandermonde_det}, while the odd case $\MinEnt^{\odd}$ turned out to be more challenging and we had to resort to numerics.
To extract the logarithmic contribution, we look at the behavior of the entropy difference $\MinEnt^{\even} - \MinEnt^{\odd}$ which we assume would capture the contribution of the moving spinon.
The full steps of the calculation are illustrated in Appendix~\ref{app:computation_of_minent_at_delta_zero}.
The evaluation of the entropy difference, as reported in Fig.~\ref{fig:minent_delta_zero}, shows that it behaves as
\begin{equation}
    \eval{\MinEnt^{\odd} - \MinEnt^{\even}}_{\Delta = 0} \simeq \frac{1}{4} \log L + c_0
    \label{eq:min_entropy_difference_delta_zero}
\end{equation}
with very good accuracy, where $c_0$ is a constant specific to $\Delta = 0$.
The disappearance of the linear term in the difference \eqref{eq:min_entropy_difference_delta_zero} can be explained by the fact that the entropy per site must be the same in the thermodynamic limit, independently of the parity.
This requirement, however, still allows corrections of the form $\log L$, which, when divided by $L$, vanishes in the limit $L \to \infty$.


\paragraph{Critical phase and iMPS}
So far we have only stated that the log term coefficient at $\Delta = 0$ is $1/4$ while at $\Delta = 1/2$ is $1/3$.
Nonetheless, we can integrate these results into a much more general statement with the help of \emph{infinite Matrix Product States (iMPS) Ansatz} \cite{cirac2010ansatz, herwerth2015spinstates, manna2018chain}.
In short, the iMPS Ansatz for the critical phases of spin chains substitutes the finite-dimensional tensors at each site with vertex operators of a chiral boson field, effectively working with an infinite bond dimension.
Therefore, on a chain with $L$ sites, the different components of the ground state are given by the expectation values of $L$ vertex operators, where the charges of the operators depend on the spin states and $\Delta$.
Following \cite{cirac2010ansatz}, we have the following Ansatz:
\begin{equation}
    \Psi_{\text{iMPS}}^{\alpha}(n_1 \dots n_N) \propto
    e^{i \pi \sum_{j} n_j } \prod_{i > j}^{N} \qty(e^{i 2 \pi n_i / L} - e^{i 2 \pi n_j / L})^{4 \alpha},
    \label{eq:IMPS}
\end{equation}
where the $n_j$'s are the position of the spin ups and $\alpha \in [0, 1/2]$ is the variational parameter of the Ansatz
\footnote{
Due to the charge neutrality condition, in the case of an odd chain, we have to put an extra vertex operator at infinity to have a non-vanishing wavefunction, which in the end contributes with only a global phase factor \cite{herwerth2015spinstates}.
}.
The normalization constant of \eqref{eq:IMPS} is
\begin{equation}
    Z_{\alpha}(L,N) = \sum_{\{n_j\}} \prod_{n_i > n_j}^{N}
    \abs{e^{i 2 \pi n_i / L} - e^{i 2 \pi n_j / L}}^{8 \alpha},
    \label{eq:imps_normalization}
\end{equation}
where the sum is over all the possible position configurations of $1 \leq n_1 < \dots < n_N \leq L$.

The Ansatz \eqref{eq:IMPS} has a fidelity of over $99 \%$ for chains up to 20 sites and the relationship between $\Delta$ and $\alpha$ is given by \cite{cirac2010ansatz}
\begin{equation}
    \Delta = - \cos (2 \pi \alpha).
    \label{eq:delta_vs_alpha}
\end{equation}
Notice that at $\Delta = 0$ and $1/2$ we have $\alpha = 1/4$ and $1/3$, which are exactly the values of the log term coefficient in \eqref{eq:min_entropy_difference_delta_zero} and \eqref{eq:minent_delta_half}, respectively.
Therefore, assuming that $\MinEnt^{\odd}$ (for $L$ odd) scales as $a L + b \log L + c$, \emph{we conjecture that $b$ is the same parameter $\alpha$ in \eqref{eq:IMPS} that satisfies \eqref{eq:delta_vs_alpha}}.
In terms of the radius $R$ and Luttinger parameter $K$, it is equivalent to stating that
\begin{equation}
    b = \alpha = \pi R^2 = \frac{1}{4 K}.
    \label{eq:b_vs_R_vs_K}
\end{equation}

This conjecture can be further justified by comparing the Ansatz \eqref{eq:IMPS} with the $\Delta = 0$ case \eqref{eq:slater_vandermonde_det}, in the following way.
Suppose the maximal probability at any $\Delta$ is given by \eqref{eq:IMPS}, that is $\pmax(L,N)_{\Delta} = \abs{\Psi_{\text{iMPS}}^{\alpha}(n_j)}^2$ with configuration $n_j = 2j - 1$ (where $j = 1, \dots, N$) and the corresponding $\alpha$ in \eqref{eq:delta_vs_alpha}.
We define
\begin{equation}
    F(L, N) = \prod_{j > k}^{N} \abs{ e^{i 4 \pi j / L} - e^{i 4 \pi k / L} }^2
    \label{eq:Phi_def}
\end{equation}
such that
\begin{equation}
    \pmax(L, N)_{\Delta} = \frac{F(L, N)^{4 \alpha}}{Z_{\alpha}(L, N)}
    \label{eq:pmax_and_Phi}
\end{equation}
We recognize \eqref{eq:Phi_def} as the squared Slater determinant \eqref{eq:slater_vandermonde_det} for the maximal state at $\Delta = 0$, minus a factor $L^N$.
Hence we can write
\begin{equation}
    F(L, N) = L^N \pmax(L, N)_{\Delta = 0}
    \label{eq:F_and_pmax_delta_zero}
\end{equation}
Now consider the entropy difference
\begin{equation*}
    \eval{\MinEnt^{\odd} - \MinEnt^{\even}}_{\Delta} =
    - \log \frac{\pmax^{\odd}(N)_{\Delta}}{\pmax^{\even}(N)_{\Delta}}.
\end{equation*}
Inserting \eqref{eq:pmax_and_Phi} and then \eqref{eq:F_and_pmax_delta_zero} in the above yields
\begin{multline}
    \MinEnt^{\odd} - \MinEnt^{\even} =
    - 4 \alpha \log \frac{\pmax^{\odd}(N)_{\Delta = 0}}{\pmax^{\even}(N)_{\Delta = 0}} - \\
    - 4 \alpha N \log \qty( 1 + \frac{1}{2 N} )
    + \log \frac{Z^{\odd}_{\alpha}(N)}{Z^{\even}_{\alpha}(N)}.
    \label{eq:min_entropy_difference_alpha}
\end{multline}
The first term in \eqref{eq:min_entropy_difference_alpha} is the entropy difference \eqref{eq:min_entropy_difference_delta_zero} at $\Delta = 0$, which scales as $ \frac{1}{4} \log L$, multiplied by $4 \alpha$.
The second term meanwhile tends to $2 \alpha$ for $N \to \infty$.
Furthermore, for $\Delta = 0$ and $1$ ($\alpha = 1/4$ and $1/2$) the ratio $Z^{\odd}_{\alpha} / Z^{\even}_{\alpha}$ goes to a constant for $N \to \infty$, which can be proved by recognizing the discrete Coulomb gas partition function \cite{gaudin1973coulomb, haldane1988rvb} in \eqref{eq:imps_normalization} (see Appendix~\ref{app:imps_ansatz_and_the_discrete_coulomb_gas} for more details).
For this reason, we make the reasonable assumption that the ratio $Z^{\odd}_{\alpha} / Z^{\even}_{\alpha}$ is constant in the limit $N \to \infty$ for $0 < \alpha \leq 1/2$.
Putting together all these results leads to the following statement:
\begin{equation}
    \MinEnt^{\odd} - \MinEnt^{\even} \sim \alpha \log L + \text{const},
    \label{eq:min_entropy_difference_final}
\end{equation}
which tells us that the log term coefficient is indeed the iMPS parameter $\alpha$.
From a CFT point of view, we argue that the extra log term that appears in \eqref{eq:min_entropy_difference_final} is the result of an extra primary field present in the odd ground state.
This is because the latter does not actually correspond to the CFT vacuum, as it happens for the even ground state, but it contains an extra spin-1/2 excitation \cite{alcaraz1988xxz}.
In Appendix~\ref{app:cft_argument} we propose a heuristic argument in favor of this point of view.

For this reason, we can regard the coefficient in front of the log term as universal, because only the low-energy degrees of freedom are involved, which are described by an appropriate CFT.
This explains the dependency of the log term coefficient on the compactification radius $R$ or, equivalently, the Luttinger parameter $K$.
This connection is made even more evident by the use of the iMPS Ansatz.
The physical interpretation of the identification between the log term coefficient and the variational parameter $\alpha$ of the iMPS is that $\alpha$ governs the power-law behavior of the spin-spin correlations in the iMPS, which is precisely the role played by the Luttinger parameter $K$ \cite{cirac2010ansatz}.

In order to corroborate \eqref{eq:min_entropy_difference_final} we have computed $\MinEnt$ for $L=7,9,\dots,23$ with exact diagonalization in the range $\Delta \in (-1, 1]$.
At each $\Delta$ then, we have fitted the data against $a L + b \log L + c$ and extracted a value of $b(\Delta)$.
For comparison, we repeated the same procedure but this time computing $\MinEnt$ from the iMPS Ansatz \eqref{eq:IMPS} in the same range of parameters.
The final results are shown in Fig.~\ref{fig:log_coefficient} and we see a good agreement between the two.
On top of that, both seem to match the expected curve $\Delta = - \cos(2 \pi b)$.
Due to the relatively small system sizes, any discrepancies can be attributed to finite-size effects that are still somewhat relevant.

\begin{figure}[t]
    \centering
    \includegraphics[width=0.48\textwidth]{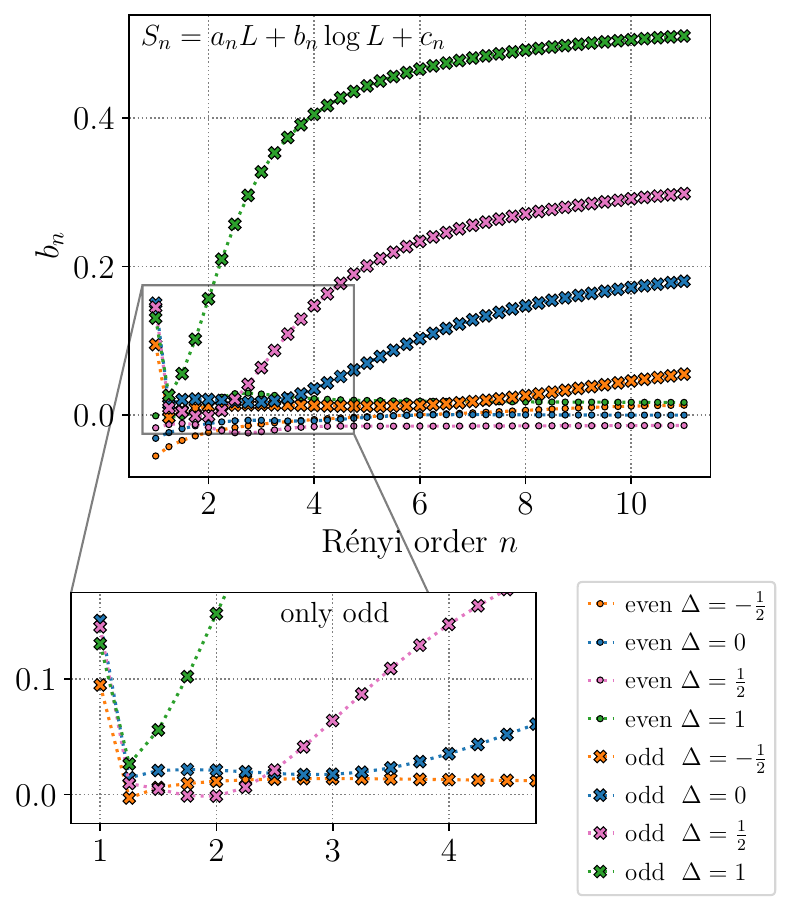}
    \caption{The log term coefficient $b_n$, as a function of $n$, in the scaling of the \Renyi{} entropy $S_n$ for some specific values of $\Delta$.
    The entropies have been obtained numerically with exact diagonalization and have been fitted against the function $S_n = a_n L + b_n \log L + c_n$.
    For the odd case the sizes $L = 11, 13, \dots, 23$ were considered, while for the even case $L = 10, 12, \dots, 20$.
    The enlarged region shows only the odd case near the von Neumann entropy limit $n \to 1$.}
    \label{fig:renyi_scaling}
\end{figure}

\paragraph{Away from $n = \infty$}
It is interesting to ask if the log term correction survives when getting away from the limit $n = \infty$.
For this reason we have computed the \Renyi{} entropy $S_n$ (with finite $n$) for some specific values of $\Delta$, both for the odd and even case, and plotted the resulting log term.
The results are shown in Fig.~\ref{fig:renyi_scaling}.

We have found that, indeed, the log term survives in the odd case when $n < \infty$, with a strength dependent on $\Delta$ and a discontinuous behaviour when $n \to 1$ (the von Neumann entropy limit).
Another observation we make is that, far away enough from $n=1$, $b_n$ appears to be an increasing function of $n$.
The small log corrections in the even case can reasonable be considered as artifacts of the relatively small system sizes.

A similar scenario have been studied in \cite{stephan2011phase} where the authors have studied the Shannon-\Renyi{} entropy of Luttinger liquids as a function of the order $n$.
They have found that the subleading constants in the entropy scaling goes through a phase transition for $n$ above a critical value $n_c$, dependent on the radius $R$, and for $n > n_c$ the entropy is regarded to be in a locked phase.
We point out that in \cite{stephan2011phase} no distinction is made between odd and even sizes.


\paragraph{Conclusion and outlooks}
In this work, we have shown that an extra term appears in the min-entropy $\MinEnt$ scaling of the spin-1/2 XXZ spin chain, only when considering \emph{odd system sizes} in the critical regime $-1 < \Delta \leq 1$.
This term is logarithmic in $L$, hence it does not disappear in the infinite volume limit $L \to \infty$.
The effect is due to an extra spin excitation present in the ground state when the size is odd, which corresponds to a spinon state, meaning that it does not have a vanishing total spin $S^z$ and is double degenerate.
We have also shown that the overall coefficient of this log term depends directly on the compactification radius $R$, or equivalently the Luttinger parameter $K$, meaning that is determined by the low-energy field theory of the critical XXZ chain.
This implies that these entanglement effects can be used to probe the spectrum of the underlying CFT.

Furthermore, numerical simulations show that indeed logarithmic corrections can still be observed in the Shannon-\Renyi{} $\MinEnt$ entropy when $n \neq \infty$.
Even though the corrections seems to display a dependence on both the index $n$ and the parameter $\Delta$, it is still unclear the nature of the relationship and if there are any universal aspect to it.

Given the correspondence between Shannon entropy in 1+1D and entanglement entropy in 2+1D, these results prove that there can be effects on the entanglement of the quantum Six-Vertex model that are not related to the manifold topology and smoothness \cite{fradkin2006entanglement}, or boundary conditions \cite{hsu2009entanglement, zaletel2011entanglement}.
Other types of effects on entanglement due to an even/odd length disparity can also be found in \cite{ju2012entanglement,stephan2013entanglement}.
Even though the authors are concerned with similar critical wavefunctions in 2+1D, they consider a cylindrical subsection of a torus instead of a half-infinite cylinder.
Then, they observe some odd behavior (oscillations in the second \Renyi{} entropy and $\MinEnt$) that depends on the parity of the \emph{length} of the cylinder.
These findings, we can say, are ``orthogonal'' to the results of this paper, because here the parity of the \emph{diameter} is considered instead.

One could consider other types of spin chains, like for example the Ising model which is the quantum Hamiltonian of a special point of the Eight-Vertex Model.
Contrary to the XXZ model, the maximal state of the Ising model is the ferromagnetic state $\ket{\uparrow \cdots \uparrow}$ (or $\ket{\downarrow \cdots \downarrow}$).
We have verified numerically that there are no oddities in this spin chain, the maximal state is always the ferromagnetic state independently of the size parity.

\begin{acknowledgments}
    SP and ER acknowledge the financial support received from the IKUR Strategy under the collaboration agreement between the Ikerbasque Foundation and UPV/EHU on behalf of the Department of Education of the Basque Government.

    ER acknowledges support from the BasQ strategy of the Department of Science, Universities, and Innovation of the Basque Government. ER is supported by the grant PID2021-126273NB-I00 funded by MCIN/AEI/ 10.13039/501100011033 and by ``ERDF A way of making Europe'' and the Basque Government through Grant No. IT1470-22.

    GS acknowledges financial support through the Spanish MINECO grant PID2021-127726NB-I00 and the CSIC Research Platform on Quantum Technologies PTI-001.

    JC is supported by the PIF 2022 grant funded by UPV/EHU.

    This work was supported by the EU via QuantERA project T-NiSQ grant PCI2022-132984 funded by MCIN/AEI/10.13039/501100011033 and by the European Union ``NextGenerationEU''/PRTR. This work has been financially supported by the Ministry of Economic Affairs and Digital Transformation of the Spanish Government through the QUANTUM ENIA project called – Quantum Spain project, and by the European Union through the Recovery, Transformation, and Resilience Plan – NextGenerationEU within the framework of the Digital Spain 2026 Agenda. This work has been partially funded by the Eric \& Wendy Schmidt Fund for Strategic Innovation through the CERN Next Generation Triggers project under grant agreement number SIF-2023-004.

    The code used for the numerical simulations and graphs, including the data for the latter, can be found at \cite{logcorrections-repo}
\end{acknowledgments}


\appendix

\section{Computation of \texorpdfstring{$\MinEnt$}{S inf} at \texorpdfstring{$\Delta = 1/2$}{Delta = 1/2}}
\label{app:computation_of_minent_at_delta_half}

In this appendix we use the results of \cite{razumov2001spin}, which have been partially proven \cite{razumov2007spin}, to compute $\MinEnt$ for odd sizes.
We note that there are some slight differences between our case and the one analyzed by the authors, that have been pointed out in \cite{Note1}.
Let $L = 2 N + 1$ be the size of the chain and consider the ground state $\ket*{\gs^{\odd}(N)}_{\Delta = 1/2}$ with total spin $S^z = +1/2$, which means $N$ spin downs and $N + 1$ spin ups.
Let $1 \leq n_1 < \dots < n_{N}$ be the positions of the spins down, and let $\psi_{n_1 \dots n_N}$ be the amplitude of the configuration $\ket{n_1 \dots n_N}$.
Then, the ground state can be written as
\begin{equation}
    \ket*{\gs^{\odd}(N)}_{\Delta = \onehalf} =
    \frac{1}{\mathcal{N}_N} \sum_{n_1 \dots n_N} \psi_{n_1 \dots n_N} \ket{n_1 \dots n_N}
\end{equation}
Due to translational invariance, all the states that differ by just a constant shift of the $\{n_j\}$ have the same amplitude.
We renormalize $\ket*{\gs^{\odd}(N)}$ such that $\psi_{1,2,\dots,N} = 1$.

Then, it is conjectured that all the coefficients $\psi$ are positive integers and $\psi_{1,3,5,\dots,2N - 1} \equiv \psi_{\text{max}}^{N}$ is the highest weight state with:
\begin{equation}
    \psi_{1,3,5,\dots,2N - 1} =
    \prod_{j = 0}^{N - 1} \frac{(3j + 1)!}{(N + j)!} =
    A(N),
    \label{eq:max_state_delta_half}
\end{equation}
where $A(N)$ is the number of alternating sign matrices of size $N \times N$.
Meanwhile, the norm $\mathcal{N}_{N}$ is conjectured to be:
\begin{equation}
    \mathcal{N}_{N} =
    \frac{\sqrt{3^{N}}}{2^{N}}
    \frac{2 \cdot 5 \cdots (3N - 1)}{1 \cdot 3 \cdots (2N - 1)} A(N).
    \label{eq:normalization_delta_half}
\end{equation}

We want to compute the asymptotic behavior of
\begin{equation}
    \begin{split}
        \pmax(2N+1, N)_{\Delta = \frac{1}{2} }
        & = \abs{ \frac{\psi_{\text{max}}^{N}}{\mathcal{N}_{N}} }^2 \\
        & = \qty[ \frac{\sqrt{3^{N}}}{2^{N}} \prod_{j = 1}^{N} \frac{(3j - 1)}{(2j -1)}   ]^{-2},
    \end{split}
    \label{eq:max_prob_delta_max}
\end{equation}
where we have done just a simple substitution of \eqref{eq:max_state_delta_half} and \eqref{eq:normalization_delta_half}.
Using the identity
\begin{equation*}
    \Gamma\qty(n + \frac{p}{q} ) =
    \frac{1}{q^n} \Gamma \qty( \frac{p}{q} ) \prod_{j = 1}^{n} (qj - q + p),
\end{equation*}
we can rewrite the product series as
\begin{equation*}
    \prod_{j = 1}^{N} \frac{(3j - 1)}{(2j - 1)} =
    \qty(\frac{3}{2})^{\!\!N} \, \frac{\Gamma(1/2)}{\Gamma(2/3)} \, \frac{\Gamma(N + 2/3)}{\Gamma(N + 1/2)}
\end{equation*}
We substitute the above expression and $N = (L-1)/2$ into \eqref{eq:max_prob_delta_max} and obtain
\begin{equation*}
    \pmax^{-1} =
    \qty( \frac{3 \sqrt{3}}{2} )^{L - 1}
    \qty( \frac{\Gamma(1/2)}{\Gamma(2/3)} )^2
    \qty( \frac{\Gamma(L/2 + 1/6)}{\Gamma(L/2)} )^2
\end{equation*}
As a last step, we use the asymptotic behavior $\Gamma(x + \alpha) \sim \Gamma(x) x^{\alpha}$, which shows that the last term behaves like $(L/2)^{1/3}$.
We also use the identity $\Gamma(1/2) = \sqrt{\pi}$. Putting everything together, we finally arrive at an expression of the min-entropy in the large $L$ limit:
\begin{equation}
    \MinEnt(L) = L \log ( \frac{3 \sqrt{3}}{4} )
    + \frac{1}{3} \log \qty( \frac{L}{2} )
    + \log ( \frac{4 \pi}{3 \sqrt{3} \Gamma(2/3)^2} ).
    \label{eq:minent_delta_half_theor}
\end{equation}

We see that there is an extra $\log$ term in the min-entropy:
\begin{equation}
    S_{\log} = \frac{1}{3} \log L
\end{equation}


\section{Computation of \texorpdfstring{$\MinEnt$}{S inf} at \texorpdfstring{$\Delta = 0$}{Delta = 0}}
\label{app:computation_of_minent_at_delta_zero}

The amplitudes of the ground state of the XX ($\Delta = 0$) chain in the occupational basis, with system size $L$ and $N$ particles, is given by the Slater determinant
\begin{equation}
    \braket{n_1 \dots n_N}{\gs}_{\Delta = 0} = \det \mathsf{M}
\end{equation}
with
\begin{equation}
    \mathsf{M} = \qty(\frac{e^{i q_j n_k}}{\sqrt{L}})_{j,k=1}^{N}
    \label{eq:slater_det_app}
\end{equation}
where $q_j$ are the occupied momenta in the ground state and $n_k$ are the particle positions.
The former are all equally spaced, $\Delta q = 2 \pi /L$, and the maximal probability is reached when $n_j = 2j - 1$ with $j = 1, \dots, N$ (up to a translation).
As mentioned before, for $L$ even we have $N = L/2$ while for $L$ odd $N$ can be either $(L-1)/2$ or $(L+1)/2$.
We write $\pmax = \abs{\braket{i_{\text{max}}}{\gs}}^2$ as
\begin{equation}
    \pmax(L, N)_{\Delta = 0}
    = \frac{1}{L^N}  \abs{\det \mathsf{V}}^2
    = \frac{1}{L^N}  \det ( \mathsf{V}^{\dagger} \mathsf{V})
    \label{eq:maxprob_delta_zero}
\end{equation}
with
\begin{equation}
    \mathsf{V}(L, N) = \qty(e^{i 4 \pi n m / L})_{n,m=1}^{N},
\end{equation}
where global phase factors in \eqref{eq:slater_det_app} have been removed, as they do not contribute to the modulus.
Notice that it is a Vandermonde matrix and its determinant can be written as
\begin{equation}
    \det \mathsf{V}(L, N) = \prod_{n > m}^{N} \qty( e^{i 4 \pi n / L} - e^{i 4 \pi m / L} ).
\end{equation}
It is clear that $\abs{\det \mathsf{V}(L, N)}^2 = F(L, N)$, where the latter is defined in \eqref{eq:Phi_def}.

We can compute $\det \mathsf{V}$ exactly in the even case $L = 2N$, where it reduces to a Fourier matrix:
\begin{equation*}
    \mathsf{V}_{\even} = \qty( e^{i 2 \pi n m / N} )_{n,m=1}^{N}.
\end{equation*}
The Hadamard inequality states that, given a $N \times N$ matrix $\mathsf{A}$ with column vectors $\bm{w}_j$, the following holds:
\begin{equation}
    \abs{\det \mathsf{A}} \leq \prod_{j}  \norm{\bm{w}_j},
\end{equation}
where $\norm{\cdot}$ is the Euclidean norm.
In the case of non-zero vectors, the equality is saturated only if the $\bm{w}_j$ are orthonormal.
For $\mathsf{V}^{\even}$, it can be easily verified that the columns $\bm{v}_j = \qty( e^{i 2 \pi n j/N} )$ are orthogonal with norm $\norm{\bm{v}_j} = \sqrt{N}$.
Therefore
\begin{equation*}
    \det \mathsf{V}_{\even} = N^{N/2},
\end{equation*}
which means that
\begin{equation*}
    \pmax^{\even}(N)_{\Delta = 0} = \frac{N^{N}}{(2N)^N} = \frac{1}{2^N}
\end{equation*}
and consequently $\MinEnt^{\even} = N \log 2$.

On the other hand, the odd case is not so simple.
The number of particles is fixed but the size is increased by one.
Then, the matrix
\begin{equation*}
    \mathsf{V}_{\odd} = \qty( e^{i 2 \pi n m / (N + 1/2)} )_{n, m = 1}^N
\end{equation*}
has no longer orthogonal columns, therefore the Hadamard inequality is strict.
This already gives us a hint about the presence of some extra term in the min-entropy.
To isolate and quantify these extra terms we look at the ratio
\begin{equation}
        \frac{\pmax^{\odd}(N)_{\Delta=0}}{\pmax^{\even}(N)_{\Delta = 0}}
        = \frac{(2N)^N}{(2N + 1)^N} \frac{\det (V_{\odd}^{\dagger} V_{\odd})}{\det (V_{\even}^{\dagger} V_{\even})},
    \label{eq:maxprob_ratio_delta_zero}
\end{equation}
where we have used \eqref{eq:maxprob_delta_zero}.
The prefactor in \eqref{eq:maxprob_ratio_delta_zero} tends to $e^{-1/2}$ for $N \to \infty$, while with some algebra it can be shown that the ratio of determinants can be written as $\det \mathsf{W}$ with
\begin{equation}
   \mathsf{W}_{nm} =
   \begin{cases}
       \displaystyle \frac{1}{N(1 + \epsilon^{m - n})} & n \neq m \\
       1 & n = m
   \end{cases}
\end{equation}
where $\epsilon = e^{i 2 \pi /(2N + 1)}$.

By taking the $- \log$ of \eqref{eq:maxprob_ratio_delta_zero} we have
\begin{equation*}
    \MinEnt^{\odd} - \MinEnt^{\even} =
    - \log \det \mathsf{W} + \frac{1}{2}
\end{equation*}
To evaluate the dependence of this quantity with the system size, we resorted to numerical methods and fitted $- \log \det \mathsf{W}$ against the function $a N + b \log N + c$, as seen in Fig.~\ref{fig:minent_delta_zero} (see the caption for obtained values of $a$, $b$ and $c$). From the results of the fits, we gather that $b \simeq 1/4$ with good confidence.
Thus, we can conclude that
\begin{equation*}
    \MinEnt^{\odd} - \MinEnt^{\even} = \frac{1}{4} \log N + \text{const}.
\end{equation*}


\section{iMPS Ansatz and the discrete Coulomb gas}
\label{app:imps_ansatz_and_the_discrete_coulomb_gas}

The discrete Coulomb gas is a classical model of $N$ particles on a circle with $L$ sites and interaction potential
\begin{equation*}
    W(n_1, \dots, n_N) = - \sum_{j < k}^{N} \log \abs{\varepsilon^{n_j} - \varepsilon^{n_k}},
    \quad \varepsilon = e^{i 2 \pi / L}.
\end{equation*}
introduced by \citeauthor{gaudin1973coulomb} in \cite{gaudin1973coulomb}. The Boltzmann statistics are given by the weights
\begin{equation}
    e^{- \beta W(n_1 \dots n_N)} =
    \prod_{j < k}^{N} \abs{\varepsilon^{n_j} - \varepsilon^{n_k}}^{\beta},
\end{equation}
where $\beta$ is the inverse temperature (or ``coolness'' \cite{baez2024entropy}). When compared with \eqref{eq:IMPS}, it has the same form of $\abs{\Psi_{\text{iMPS}}}^2$ with the correspondence
\begin{equation}
    \beta = 8 \alpha.
\end{equation}
The partition function of the discrete Coulomb gas, defined as
\begin{equation}
    Q_{\beta}(L, N) =
    \frac{1}{L^{N \beta / 2}}
    \sum_{1 \leq n_1 < \dots < n_N \leq L} e^{- \beta W(n_1 \dots n_N)},
\end{equation}
was first computed by \citeauthor{gaudin1973coulomb} for some special values of $\beta$. We can use the partition function $Q(L, N, \beta)$ to express $Z(L, N, \alpha)$ in \eqref{eq:imps_normalization}
as
\begin{equation}
    Z_{\alpha}(L, N) = L^{4 N \alpha} Q_{8 \alpha}(L, N),
    \label{eq:norm_const_and_part_func}
\end{equation}
thus we can reuse the results for the Coulomb gas for our purposes.

\citeauthor{gaudin1973coulomb} computed $Q_{\beta}(L, N)$ exactly for $\beta = 0, 1, 2$, and $4$. Remember that the anisotropy $\Delta$ and $\alpha$ are related by $\Delta = - \cos(2 \pi \alpha)$, where $\alpha \in [0, 1/2]$ and $\Delta \in [-1, 1]$. Hence, we have the following table of values relating $\beta$, $\alpha$ and $\Delta$:
\begin{center}
    \begin{tabular}{ c c c}
        \toprule
        ~~$\beta$~~ & ~~$\alpha$~~ & $\Delta$ \\
        \midrule
        $0$ & $0$ & $-1$ \\
        $1$ & $1/8$ & $- 1/ \sqrt{2}$  \\
        $2$ & $1/4$ & $0$ \\
        $4$ & $1/2$ & $1$ \\
        \bottomrule
    \end{tabular}
\end{center}
We ignore the point $\beta = 0$ ($\Delta = -1$), as it is not critical, and $\beta = 1$ ($\Delta = - 1/ \sqrt{2}$) due to the cumbersome form of $Q_1(L, N)$. For $\beta = 2$ we have
\begin{equation}
    Q_2(L, N) = 1,
    \label{eq:part_func_beta_2}
\end{equation}
which confirms that $Z_{1/4}(L, N) = L^N$. Moreover, for $\beta = 4$ we have
\begin{equation}
    Q_4(L, N) =
    \begin{cases}\displaystyle
        \displaystyle
        \frac{(2N)!}{2^N N!} \frac{1}{L^N} & \text{if $2N \leq L$} \\[10pt]
        \displaystyle
        \frac{(2L - 2N)!}{2^N (L - N)!} \frac{1}{L^{L-N}} & \text{if $2N \geq L$}
    \end{cases}
    \label{eq:part_func_beta_4}
\end{equation}

In \eqref{eq:min_entropy_difference_alpha} we have argued that the scaling of the entropy difference $\MinEnt^{\odd} - \MinEnt^{\even}$ depends only on
\begin{equation*}
    - 4 \alpha \log \frac{F^{\odd}(N)}{F^{\even}(N)},
\end{equation*}
which shows a logarithmic behavior. For this to be true, one has to show that the ratio
\begin{equation}
    x = \frac{Z_{\alpha}^{\odd}(N)}{Z_{\alpha}^{\even}(N)}
    = \frac{(2N+1)^{4N \alpha} Q_{8 \alpha}(2N + 1, N)}{(2N)^{4 N \alpha} Q_{8 \alpha}(2N, N)}
\end{equation}
asymptotically tends to a constant contribution to the entropy difference.
This is true for $\Delta = 0$ and $\Delta = 1$ at least. In fact, recalling \eqref{eq:norm_const_and_part_func}, both for $\Delta = 0$ using \eqref{eq:part_func_beta_2} (where $\alpha = 1/4$) and for $\Delta = 1$ using \eqref{eq:part_func_beta_4} (where $\alpha = 1/2$) we have
\begin{equation}
    x = \qty( \frac{2N + 1}{2N}  )^N \!\!\!\longrightarrow e^{1/2}
    \quad \text{for } \quad
    N \to \infty.
\end{equation}
which supports our assumptions about the scaling of $\MinEnt^{\odd} - \MinEnt^{\even}$.


\section{CFT argument}
\label{app:cft_argument}

It is possible to offer a heuristic argument on CFT grounds for the presence of the logarithmic term.
Shortly, it is a scaling factor introduced by the spinon excitation in the odd ground state and the coefficient is given by the scaling dimension of the primary field that creates/annihilates spin-1/2 excitations.

In order, to compute $p_{\text{max}} = \abs{\braket{\imax}{\gs}}^2$ we need to find a representation of both $\ket{\gs}$ and $\ket{\imax}$ in terms of CFT states, for both $L$ even and odd.
For $L$ even the total spin $S^z$ is zero and the ground state $\ket*{\gseven}$ corresponds to the CFT vacuum $\ket{0}$.
Meanwhile, the maximal state is a \Neel{} state $\ket*{\neeleven}$ for which a description in terms of field states may be very difficult.
It is some high-energy state and for now, we just assume that is some descendent state with total spin $S^z = 0$.

For $L$ odd, the ground state $\ket*{\gsodd}$ is no longer the vacuum $\ket{0}$ as it has total spin $S^z = \pm 1/2$ and it is degenerate.
Instead, it corresponds to a spin-$\onehalf$ excitation which has conformal weight $\Delta_{1/2}$ \cite{eggert1992heisenberg, alcaraz1988xxz}.
Thus, we can identify the odd ground state with a primary field $\Phi$, that generates said excitation.
The state $\ket{\imax}$ will be again some high-energy state $\ket*{\neelodd}$ that represents the \Neel{} state with odd size (or a linear superposition of them).

\newcommand{\cyl}{\text{(cyl)}}
\newcommand{\plane}{\text{(pl)}}
Given the PBC on the chain, the corresponding CFT is defined on a cylinder of diameter $L$.
We parametrize the cylinder with a complex coordinate that is periodic along the real axis, i.e.~$w \equiv w + L$.
The odd ground state $\ket*{\gsodd}$ has to be defined in the ``remote past'' $ t \to - \infty$, which in complex coordinates means $w \to - i \infty$.
Therefore
\begin{equation}
    \ket*{\gsodd} = \lim_{w \to -i \infty} \Phi_{\cyl}(w) \ket{0},
\end{equation}
where $\Phi_{\cyl}$ is the primary field on the cylinder.
We suppose it is chiral, meaning that it depends only on $w$ and not on the conjugate $\overline{w}$.

The task is to evaluate the ratio
\begin{equation}
    \begin{split}
        \frac{\pmax^{\odd}}{\pmax^{\even}}
        & = \frac{
            \abs*{\braket*{\neelodd}{\gsodd}}^2
        }{
            \abs*{\braket*{\neeleven}{\gseven}}^2
        } \\
        & = \lim_{w \to -i \infty} \frac{
            \abs*{\mel*{\neelodd}{\Phi_{\cyl}}{0}}^2
        }{
            \abs*{\braket*{\neeleven}{0}}^2
        }
    \end{split}
\end{equation}
To evaluate the inner products that enter the ratio, we use radial quantization, they have to be evaluated on the complex plane (or Riemann sphere).
We map the cylinder onto the complex plane with the conformal transformation
\begin{equation}
    w \mapsto z = \exp \qty( - \frac{i 2 \pi w}{L}  ).
\end{equation}
In this way the remote past is mapped onto the point $z = 0$.
Using the fact that $\Phi$ is primary with weight $\Delta_{1/2}$, we find
$\Phi_{\plane}(z) = (- i 2 \pi z / L)^{- \Delta_{1/2}} \Phi_{\cyl}(w)$.
Thus the odd ground state in radial quantization is given by
\begin{equation}
    \ket*{\gsodd} = \qty( - \frac{i 2 \pi}{L} )^{\Delta_{1/2}} \lim_{z \to 0} z^{\Delta_{1/2}} \Phi_{\plane}(z) \ket{0}
    \label{eq:gsodd_plane}
\end{equation}

The extra scaling factor $z^{\Delta_{1/2}}$ cancels out when taking the Hermitian conjugate.
Hermitian conjugation on the plane corresponds to the map $z \mapsto 1/z^{\ast}$ (with $z^{\ast} = \overline{z}$ and up to a spin-dependent phase \cite{ginsparg1988cft, difrancesco2012conformal}).
Thus, we have that $\bra*{\gsodd} = [\ket*{\gsodd}]^{\dagger}$ is equal to
\begin{equation}
    \begin{split}
        \bra*{\gsodd}
        & = \qty[ \qty( - \frac{i 2 \pi}{L} )^{\Delta_{1/2}} z^{\Delta_{1/2}} \Phi_{\plane}(z) \ket{0} ]^{\dagger} \\
        & = \qty( \frac{i 2 \pi}{L} )^{\Delta_{1/2}} \bra{0} \frac{1}{\overline{z}^{\Delta_{1/2}}} \Phi_{\plane}^{\dagger}(z)
    \end{split}
    \label{eq:gsodd_plane_hc}
\end{equation}
with $z \to 0$.
Using \eqref{eq:gsodd_plane} and \eqref{eq:gsodd_plane_hc} we have
\begin{multline}
    \pmax^{\odd} = \qty(\frac{2 \pi}{L})^{2 \Delta_{1/2}}
    \lim_{z \to 0} \mel*{\neelodd}{\Phi_{\plane}(z)}{0} \times \\
    \times \mel*{0}{\Phi_{\plane}^{\dagger}}{\neelodd}.
    \label{eq:pmax_odd}
\end{multline}
We have discarded the factor $(z / \overline{z})^{\Delta_{1/2}}$ because it is at most a phase and $\pmax^{\odd}$ has to be a real number.

Given that $\Phi_{\plane}$ is the primary field that creates a spin-1/2 excitation, then it is reasonable to say that $\Phi_{\plane}^{\dagger}$ annihilates a spin-1/2 excitation.
Therefore, we can state heuristically that
\begin{equation}
    \lim_{z \to 0} \Phi_{\plane}^{\dagger}(z) \ket*{\neelodd} \approx \ket*{\neeleven},
    \label{eq:phi_on_neelodd}
\end{equation}
meaning that if we start with $\ket*{\neelodd}$ and then apply $\Phi_{\plane}$ to it, we remove the extra spin excitation, thus obtaining something similar to $\ket*{\neeleven}$.
Conversely, $\bra*{\neelodd} \Phi_{\plane} \approx \bra*{\neeleven}$ as it is just the Hermitian conjugate of the above expression.
A precise evaluation would require a much deeper knowledge about $\ket*{\neelodd}$ and $\ket*{\neeleven}$, which is worth investigating.
Substituting \eqref{eq:phi_on_neelodd} in \eqref{eq:pmax_odd} yields
\begin{equation}
    \pmax^{\odd} \simeq \qty( \frac{2 \pi}{L} )^{2 \Delta_{1/2}} \pmax^{\even},
\end{equation}
therefore
\begin{equation}
    \MinEnt^{\odd} - \MinEnt^{\even} \simeq 2 \Delta \log L + \text{const.}
\end{equation}
which has the same form of \eqref{eq:min_entropy_difference_final}.

The last remaining step is to evaluate the exponent $\Delta_{1/2}$.
Coming from the iMPS Ansatz \cite{cirac2010ansatz}, spin states are created by the vertex operators $A_{s_i} = \chi_i : e^{i s_i \sqrt{\alpha} \varphi(z)} :$ of a chiral boson field $\varphi(z)$.
They have conformal weight $\alpha / 2$, which in terms of compactification radius means $\pi R^2 / 2$.
If we take $\Phi$ as one of these vertex operator, then $\Delta_{1/2} = \alpha / 2$ and \eqref{eq:b_vs_R_vs_K} is proven.
We see also from \cite{eggert1992heisenberg, alcaraz1988xxz} that the conformal weight $\Delta_{1/2}$ is equal to $ \pi R^2 / 2$, again following our findings.


\bibliography{biblio}

\end{document}